%% file: paper.tex
\documentclass[twocolumn,hyperpdf,amsmath,amssymb,aps,prl,10pt,superscriptaddress,nofootinbib,noeprint,preprintnumbers,floatfix]{revtex4-2}

\usepackage{graphicx, color}
\usepackage[dvipsnames]{xcolor}

\usepackage[letterspace=-10]{microtype} 

\usepackage{bm, amsmath, amsfonts, amssymb,xfrac}
\usepackage{multirow, tabularx, dcolumn}
\usepackage{mathtools, leftidx, braket, slashed, cancel, bigdelim}
\usepackage{blkarray}
\usepackage[figures]{rotating}

\usepackage{tikz}
\usepackage{soul}
\usetikzlibrary[plotmarks]
\usepackage{anyfontsize}

\usepackage[utf8]{inputenc} 
\usepackage{hyperref}
\definecolor{jlab_red}{RGB}{192,39,45}
\definecolor{jlab_orange}{RGB}{249,102,0}
\definecolor{jlab_blue}{RGB}{47,122,121}
\definecolor{jlab_green}{RGB}{65,125,10}
\definecolor{jlab_gray}{gray}{0.6}
\definecolor{magenta}{rgb}{0.5, 0, 0.5}

\newcommand\bef{\begin{figure}}
\newcommand\eef[1]{\label{fg:#1}\end{figure}}
\newcommand\beq{\begin{equation}}
\newcommand\eeq[1]{\label{#1}\end{equation}}
\newcommand\beqa{\begin{eqnarray}}
\newcommand\eeqa[1]{\label{#1}\end{eqnarray}}
\newcommand\bet{\begin{table}}
\newcommand\eet[1]{\label{tb:#1}\end{table}}
\newcommand\fgn[1]{Figure \ref{fg:#1}} 
\newcommand\eqn[1]{Eq.\ (\ref{#1})}
\newcommand\tbn[1]{Table \ref{tb:#1}} 

\hypersetup{%
pdftitle = {title},
pdfsubject = {},
pdfkeywords = {},
colorlinks = {true},
filecolor = {black},
linkcolor = {jlab_blue},
menucolor = {black},
citecolor = {jlab_green},
urlcolor = {jlab_green},
}{}

\begin{document}

%
\title{Bound isoscalar axial-vector $bc\bar u\bar d$ tetraquark $T_{bc}$ from lattice QCD using two-meson and diquark-antidiquark variational basis}
\author{M. Padmanath}
\email{padmanath@imsc.res.in}
\affiliation{The Institute of Mathematical Sciences, a CI of Homi Bhabha National Institute, Chennai, 600113, India}
\author{Archana Radhakrishnan}
\email{archana.radhakrishnan@tifr.res.in}
\affiliation{Department of Theoretical Physics, Tata Institute of Fundamental Research, \\ Homi Bhabha Road, Mumbai 400005, India }

\author{Nilmani Mathur}
\email{nilmani@theory.tifr.res.in}
\affiliation{Department of Theoretical Physics, Tata Institute of Fundamental Research, \\ Homi Bhabha Road, Mumbai 400005, India }

\preprint{IMSc/23/05, TIFR/TH/23-14}

\date{\today}
\begin{abstract}
We report a lattice QCD study of the heavy-light meson-meson interactions with an explicitly exotic flavor content 
$bc\bar u\bar d$, isospin $I\!=\!0$, and axialvector $J^P=1^+$ quantum numbers in search of possible tetraquark bound states. 
The calculation is performed at four values of lattice spacing, ranging $\sim$0.058 to $\sim$0.12 fm, and at five 
different values of valence light quark mass $m_{u/d}$, corresponding to pseudoscalar meson mass $M_{ps}$ of about 
0.5, 0.6, 0.7, 1.0, and 3.0 GeV. The energy eigenvalues in the finite-volume are determined through a variational 
procedure applied to correlation matrices built out of two-meson interpolating operators as well as  
diquark-antidiquark operators. The continuum limit estimates for $D\bar B^*$ elastic $S$-wave scattering amplitude are
extracted from the lowest finite-volume eigenenergies, corresponding to the ground states, using amplitude 
parametrizations supplemented by a lattice spacing dependence. Light quark mass $m_{u/d}$ dependence of the 
$D\bar B^*$ scattering length ($a_0$) suggests that at the physical pion mass $a_0^{phys} = +0.57(^{+4}_{-5})(17)$ fm, 
which clearly points to an attractive interaction between the $D$ and $\bar B^*$ mesons that is strong enough to 
host a real bound state $T_{bc}$, with a binding energy of $-43(_{-7}^{+6})(_{-24}^{+14})$ MeV with respect to 
the $D\bar B^*$ threshold. We also find that the strength of the binding decreases with increasing $m_{u/d}$ and 
the system becomes unbound at a critical light quark mass $m^{*}_{u/d}$ corresponding to $M^{*}_{ps} = 2.73(21)(19)$ GeV. 
\end{abstract}

\maketitle

The discovery of a doubly charmed tetraquark\footnote{We follow the nomenclature that a ``tetraquark"
refers to any bound state or resonance with dominant four-quark Fock component, whether it is 
a compact four-quark object or a two-meson molecule or a mixture of both.}, $T_{cc}$, marks an 
important milestone \cite{LHCb:2021vvq} in spectroscopy of hadrons. Phenomenologically, doubly heavy 
tetraquarks in the heavy quark limit are long hypothesized to form deeply bound states \cite{Ader:1981db,
Ballot:1983iv,Zouzou:1986qh,Heller:1986bt,Carlson:1987hh,Manohar:1992nd,Janc:2004qn,Ebert:2007rn,
Navarra:2007yw,Eichten:2017ffp,Karliner:2017qjm} with binding energy $\mathcal{O}$(100 MeV) with respect 
to the elastic strong decay threshold. While doubly bottom tetraquarks are suitable candidates for such 
deeply bound states, as predicted by multiple lattice QCD calculations \cite{Bicudo:2015kna,Francis:2016hui,
Bicudo:2017szl,Junnarkar:2018twb,Leskovec:2019ioa,Hudspith:2023loy}, $T_{cc}$ is found to be $360$ keV 
below the lowest two-meson threshold ($D^0D^{*+}$). A handful of recent experimental 
developments involving multiple heavy quark production such as the recent discoveries of $\Xi_{cc}$
\cite{LHCb:2017iph}, $T_{cc}$ \cite{LHCb:2021vvq}, reports of tri-$J/\psi$ \cite{CMS:2021qsn}, associated 
$J/\psi\Upsilon$ \cite{LHCb:2023qgu}, and di-$\Upsilon$ \cite{CMS:2016liw} productions, and recent 
proposals of inclusive search strategies \cite{Gershon:2018gda,Qin:2021zqx} augment promising prospects 
for the doubly heavy hadron sector in the near future. In light of these advancements, a doubly heavy 
tetraquark with a bottom and a charm quark with a valence quark configuration $T_{bc} \equiv bc\bar u\bar d$ 
is going to be one of the most sought-after hadron in this decade \cite{Polyakov:2023had}. In this work, 
using lattice QCD calculations, we show a clear evidence of an attractive interaction between the $D$ and 
$\bar B^*$ mesons that is strong enough to host a real bound state $T_{bc}$. This finding will further boost 
the search for such bottom-charm tetraquarks.

The phenomenological picture on deeply bound doubly heavy tetraquarks is based on a compact heavy
diquark-light antidiquark interpretation \cite{Francis:2016hui,Czarnecki:2017vco}, whereas the shallow 
binding energy of $T_{cc}$ could possibly be a reflection of its dominant noncompact molecular nature 
\cite{Janc:2004qn,Agaev:2022ast}. Bottom-charm tetraquarks form an intermediate platform, where there 
could be complicated interplay between these pictures. A collective and refined knowledge of the low 
energy spectra in all these three doubly heavy systems ($T_{bb}$, $T_{bc}$ and $T_{cc}$) could culminate 
in a deeper understanding of strong interaction dynamics across a wide quark mass regime spanning from 
charm to the bottom quarks. The isoscalar bottom-charm tetraquarks with quantum numbers [$I(J^P) = 0(1^+)$] 
have been investigated previously both using lattice \cite{Francis:2018jyb,Hudspith:2020tdf,Meinel:2022lzo} 
and nonlattice methodologies \cite{Heller:1986bt,Carlson:1987hh,Janc:2004qn,Ebert:2007rn,Chen:2013aba,
Sakai:2017avl,Eichten:2017ffp,Karliner:2017qjm,Czarnecki:2017vco,Carames:2018tpe,Park:2018wjk,Deng:2018kly,
Yang:2019itm,Agaev:2019kkz,Lu:2020rog,Tan:2020ldi,Braaten:2020nwp}. The predictions from nonlattice 
approaches are quite scattered from being unbound to deeply bound, whereas the difference in conclusions 
from the three existing lattice QCD investigations\cite{Francis:2018jyb,Hudspith:2020tdf,Meinel:2022lzo} 
call for more detailed efforts in this regard. 

In this work, we perform a lattice QCD simulation of coupled $D\bar B^*$ and $\bar BD^*$ two-meson 
channels\footnote{We work in the isosymmetric limit with no QED effects and $m_u=m_d$. Hence we choose 
to call the degenerate ($D^+B^-,~D^0\bar{B}^0$) threshold as $D\bar B$, and equivalently for others like 
$D\bar B^*$, $\bar BD^*$ and $D^*\bar B^*$.} that are the relevant lowest two strong decay thresholds, 
in the order of increasing energies, $E_{D\bar B^*} = M_{\bar B^*}+M_{D}$ and $E_{\bar BD^*}=M_{\bar B}+M_{D^*}$, 
where $M_{h}$ is the mass of the hadron $h$. The extracted finite-volume ground state energies are utilized 
to constrain the continuum extrapolated elastic $D\bar B^*$ scattering amplitudes following the L\"uscher's 
finite-volume prescription \cite{Luscher:1990ux,Briceno:2014oea}. The light quark mass $m_{u/d}$ dependence 
of the extracted amplitudes suggests a binding energy of $-43(^{+6}_{-7})(^{+14}_{-24})$ MeV for the 
$bc\bar u\bar d$ tetraquark pole with respect to $E_{D\bar B^*}$ at the physical point $m_{u/d}^{phys}$.

\bet[tbh]
  \begin{center}
	  \begin{tabular}{p{1.5cm}p{1.5cm}p{1.5cm}>{\hfill\arraybackslash}p{1.5cm}>{\hfill\arraybackslash}p{1.0cm}}
      \hline
   Label & Symbol & $a~[fm]$     & $N_s^3\times N_t$  & $M_{ps}^{sea}$ \\ \hline
$S_1$ & \pmb{\textcolor{red}{\tikz{\pgfsetplotmarksize{0.8ex}\pgfuseplotmark{diamond}}}} & 0.1207(11)   & $24^3\times64$ & 305 \\
$S_2$ & \pmb{\textcolor{magenta}{\tikz{\pgfsetplotmarksize{0.8ex}\pgfuseplotmark{pentagon}}}} & 0.0888(8)    & $32^3\times96$ & 312\\
$S_3$ & \pmb{\textcolor{blue}{\tikz{\pgfsetplotmarksize{0.7ex}\pgfuseplotmark{o}}}} & 0.0582(4)    & $48^3\times144$ & 319\\
$L_1$ & \pmb{\textcolor{OliveGreen}{\pgfsetplotmarksize{0.7ex}\tikz{\pgfuseplotmark{square}}}} & 0.1189(9)    & $40^3\times64$ & 217 \\   \hline
  \end{tabular}
  \end{center}
\caption{Relevant details of the lattice QCD ensembles used. The lattice spacing estimates 
are measured using the $r_1$ parameter \cite{MILC:2012znn}. $L_1$ refers to large spatial volume, 
and $S_1,~S_2$, and $S_3$ refer to small spatial volumes.}
\eet{lattice}

{\it Lattice setup}: We use four lattice QCD ensembles (see \tbn{lattice} for relevant details) with $N_f=2+1+1$ 
dynamical Highly Improved Staggered Quark (HISQ) fields generated by the MILC collaboration 
\cite{MILC:2012znn}. The charm and strange quark masses in the sea are tuned to their respective 
physical values, whereas the dynamical light quark masses correspond to sea pion masses as listed 
in \tbn{lattice}. We utilize a partially quenched setup on these configurations with valence quark 
fields up to the charm quark masses realized using an overlap fermion action as in 
Refs. \cite{Chen:2003im,xQCD:2010pnl}. We employ a nonrelativistic QCD (NRQCD) Hamiltonian 
\cite{Lepage:1992tx} for the bottom quark. Following the Fermilab prescription \cite{El-Khadra:1996wdx}, 
the bare charm \cite{Basak:2012py,Basak:2013oya} and bottom \cite{Mathur:2016hsm} quark masses on each 
ensemble are tuned using the kinetic mass of spin averaged $1S$ quarkonia 
$\{a\overline M_{kin}^{\bar QQ} = {3\over 4} aM_{kin}(V) + {1\over 4} aM_{kin}(PS)\}$ determined on 
the respective ensembles. The bare strange quark mass is set by equating the lattice estimate for 
the fictitious pseudoscalar $\bar ss$ meson mass to 688.5 MeV \cite{Chakraborty:2014aca}. 

For the valence $m_{u/d}$, we investigate five different cases: three unphysical quark masses 
[corresponding to approximate pseudoscalar meson masses $M_{ps}\sim$0.5, 0.6, and 1.0 GeV], 
the strange quark mass [$M_{ps}\sim$0.7 GeV] and the charm quark mass [$M_{ps}\sim$3.0 GeV]. 
We evaluate the finite-volume spectrum for all these five quark masses on all four ensembles, 
investigate the scattering of $D$ and $\bar B^*$ mesons in all five cases and then extract the 
$m_{u/d}$ (otherwise $M_{ps}$) dependence of the scattering parameters.

{\it Interpolators and measurements}: The finite-volume spectrum is determined from Euclidean two-point 
correlation functions $\mathcal{C}_{ij}(t)$, between interpolating operators $\mathcal{O}_{i,j}(\mathbf{x},t)$ 
with desired quantum numbers, given by
\beq
\mathcal{C}_{ij}(t) = \sum_{\mathbf{x}}\left<\mathbb{O}_j^{\dagger}(0)\mathcal{O}_i(\mathbf{x},t)\right> \approx \sum_n \mathbb Z_j^{n\dagger}Z_i^n e^{-E^nt}.
\eeq{c2pt}
Here $E^n$ is the energy of the $n^{th}$ state and $Z_i^n = \bra{0}\mathcal{O}_i\ket{n}$ is the operator-state 
overlap between the sink operator $\mathcal{O}_i$ and state $n$. We use $\mathbb{O}$ and $\mathbb Z$ to 
represent the source operator and overlaps to distinguish them from that for the sink as we follow 
a wall-source to point-sink construction in our $\mathcal{C}_{ij}$ evaluations. This is well-established 
procedure in ground state energy determination, despite the non-Hermitian setup in \eqn{c2pt} (see Refs.
\cite{Francis:2016hui,Francis:2018jyb,Mathur:2018epb,Junnarkar:2018twb,Hudspith:2020tdf,Mathur:2022ovu} for details).
We use the following set of linearly independent, yet Fierz related \cite{Padmanath:2015era}, operators, 
\beqa
\mathcal{O}_1(x) &=& [\bar u(x) \gamma_i b(x)][\bar d(x) \gamma_5 c(x)]  \nonumber \\&& - [\bar d(x) \gamma_i b(x)][\bar u(x) \gamma_5 c(x)] \nonumber\\
\mathcal{O}_2(x) &=& [\bar u(x) \gamma_5 b(x)][\bar d(x) \gamma_i c(x)]  \nonumber \\&& - [\bar d(x) \gamma_5 b(x)][\bar u(x) \gamma_i c(x)] \label{eq:mmops} \\
\mathcal{O}_3(x) &=& (\bar u(x)^T \Gamma_5 \bar d(x) - \bar d(x)^T \Gamma_5 \bar u(x))( b(x) \Gamma_i c(x)). \nonumber
\eeqa{mmops1}
$\mathcal{O}_1$ and $\mathcal{O}_2$ are two-meson operators of the type $D\bar B^*$ and $\bar BD^*$, 
respectively. $\mathcal{O}_3$ is a diquark-antidiquark type operator.
Here $\Gamma_k = C\gamma_k$ with $C=i\gamma_y\gamma_t$ being the charge conjugation matrix and 
the diquarks (antidiquarks) in the color antitriplet (triplet) representations. Other high lying 
two-meson ($D^*\bar B^*$) and three-meson ($D\bar B\pi$) interpolators are ignored in this analysis as 
they are sufficiently high in energy to have any effects on the extracted ground states. Bilocal 
two-meson interpolators with nonzero internal meson momenta are also not considered, which would 
be an important step ahead \cite{Wagner:2022bff}. We also compute two-point correlation functions 
for $\bar B$, $\bar B^*$, $D$, and $D^*$ mesons, using standard local quark bilinear interpolators 
($\bar q ~\Gamma~Q$) with spin structures $\Gamma\sim\gamma_5$ and $\gamma_i$ for pseudoscalar and 
vector quantum numbers, respectively. 

{\it Analysis}: The correlation matrices $\mathcal{C}$ evaluated for the basis in Eq. (\ref{eq:mmops}) 
are analyzed following a variational procedure \cite{Michael:1985ne} by solving the generalized eigenvalue 
problem (GEVP), $\mathcal{C}(t)v^n(t) = \lambda^n(t) \mathcal{C}(t_0)v^n(t)$. The eigenvalues in 
the large time limit represent the time evolution the low lying eigenenergies $\mathcal{E}^n$ as  
$\lim_{t\to\infty}\lambda^n(t) \sim A_ne^{-\mathcal{E}^nt}$. The corresponding eigenvectors $v^n(t)$ are 
related to the operator-state-overlaps $Z_i^n$. 

Eigenenergy extraction proceeds via fitting the eigenvalue correlators, $\lambda_{n}(t)$, 
or the ratios $R^n(t)=\lambda^n(t)/\mathcal{C}_{m_1}(t) \mathcal{C}_{m_2}(t)$, with the expected 
asymptotic exponential behaviour. Here, $\mathcal{C}_{m_i}$ is the two-point correlation function for 
the meson $m_i$. $R^n(t)$ is empirically known to efficiently mitigate correlated noise between 
the product of two single hadron correlators and the interacting correlator for the two-hadron 
system \cite{Green:2021qol}. Note that the automatic cancellation of the additive quark mass offset, 
inherent to NRQCD formulation, is an added advantage in using $R^n(t)$ for the fits. The systematics 
associated with the chosen time interval for fitting are assessed by varying the lower boundary of 
the time interval, $t_{min}$, with a fixed upper boundary, $t_{max}$, chosen considering the noise level. 
In \fgn{fitcompare}, we present a representative plot showing this $t_{min}$ dependence of the energy 
splittings ($\Delta \mathcal{E}^n$) determined from the fits to $\lambda^n(t)$ and  $R^n(t)$, respectively. 
The energy differences are evaluated from $\lambda^n(t)$ using the relation $\Delta \mathcal{E}^n = \mathcal{E}^n-M_{m_1}-M_{m_2}$, 
whereas the fits to $R^n(t)$ directly yield the respective estimates. We choose the  optimal $t_{min}$ 
values where the two different procedures found to agree asymptotically in time. We also perform 
additional checks considering an alternative quark smearing with different smearing widths to affirm 
our energy estimates, see Appendix A. Our final results are based on fitting the ratio correlators $R^n(t)$.
    
\bef[h]
\includegraphics[scale=0.5]{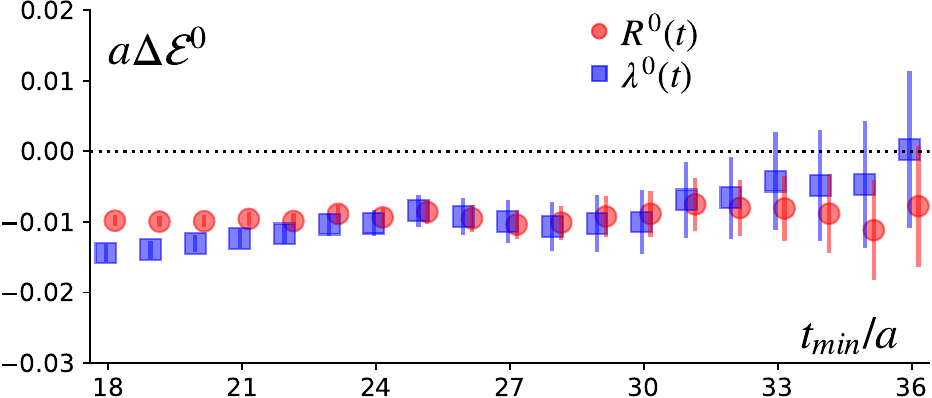}
\caption{$t_{min}$ dependence of the $\Delta \mathcal{E}^0$ fit estimates determined from the fits to $\lambda^0$
and $R^0(t)$ for the case $M_{ps} \sim 700$ MeV in the finest ensemble. Here the superscript 0 refers 
to the lowest eigenenergy. }
\eef{fitcompare}

\begin{figure}[thb!]
\includegraphics[scale=0.29]{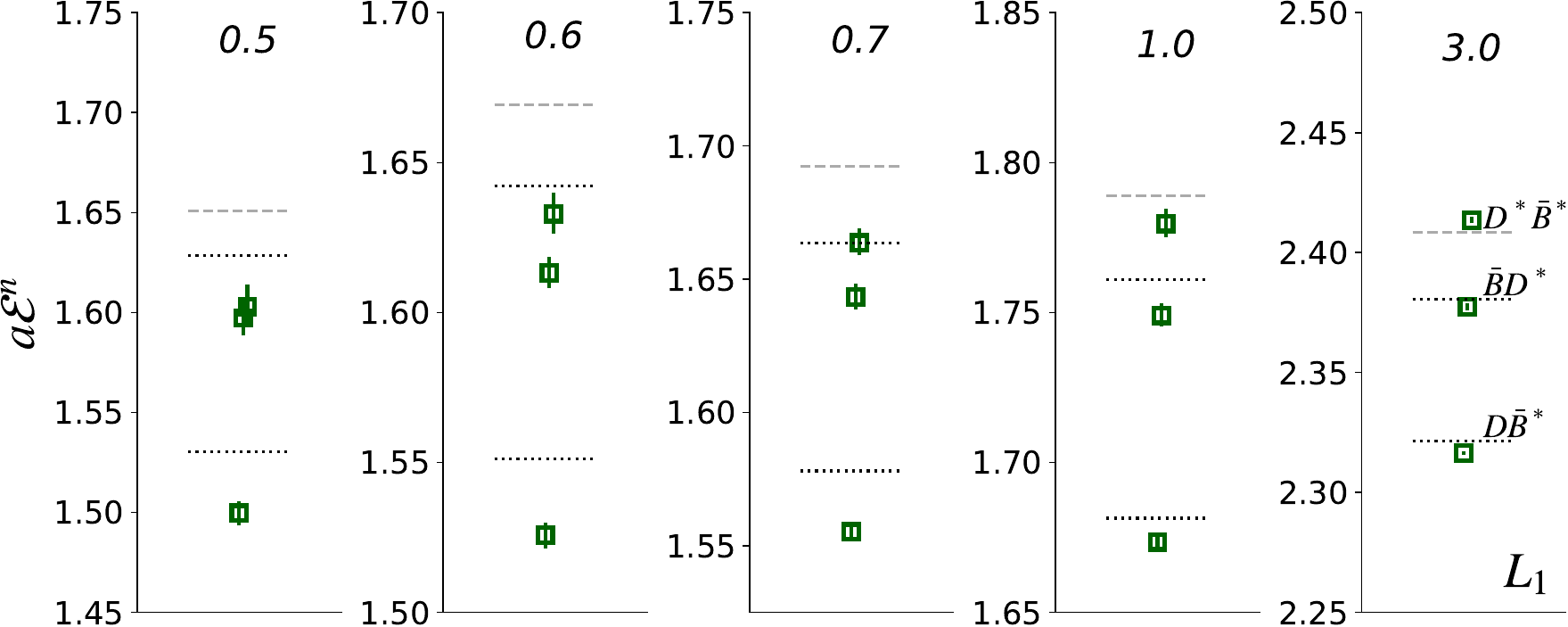} 
\caption{The GEVP eigenenergies in finite-volume for isoscalar axialvector $bc\bar u\bar d$ 
channel on the $L_1$ ensemble. Five panels show the results obtained at various pseudoscalar 
meson masses ($M_{ps}$) 0.5, 0.6, 0.7, 1.0, and 3.0, respectively.}
\eef{spectrum}
{\it Finite-volume eigenenergies}: In \fgn{spectrum}, we present the finite-volume GEVP eigenenergies, 
in lattice units, for the isoscalar axialvector $bc{\bar{u}}{\bar{d}}$ channel. The results shown 
are for the $L_1$ ensemble at the five different $m_{u/d}$ values corresponding to $M_{ps}\sim$ 0.5, 
0.6, 0.7, 1.0, and 3.0 GeV. Note that these estimates include the additive offsets related to the 
NRQCD-based bottom quark dynamics. The non-interacting two-meson energy levels corresponding to 
$D\bar B^*$ and $\bar BD^*$ thresholds are indicated as dotted horizontal line segments and those 
related to $\bar B^*D^*$ threshold by dashed lines for each $M_{ps}$. The lowest eigenenergy or 
the ground state energy is dominated by the $Z_1^0$ factor corresponding to $\mathcal{O}_1$, that 
is related to the $D\bar B^*$ threshold and is determined unambiguously by the operator $\mathcal{O}_1$, 
see Ref.~\cite{Suppl} for details. The most important observation is a clear trend for negative energy 
shifts in the ground state energies, which can be observed in all the cases, indicating a possible 
attractive interaction between the $D$ and $\bar B^*$ mesons \cite{scalarbc}. A similar pattern of 
low lying eigenenergies and ground state negative energyshifts are also observed in other ensembles, 
see details in Ref.~\cite{Suppl}. We expect that for our choice of interpolating operators and 
the accessible values of $t$, $\mathcal{E}^0$ will be an accurate estimate of $E^0$, whereas our 
setup is unable to accurately estimate excited-state energies. This means the excited eigenenergies 
presented in \fgn{spectrum} may not correspond to the higher lying elastic excitations of the $D\bar B^*$ 
channel. The location of lowest two non-interacting finite-volume levels related to the $D\bar B^*$ 
channel along with the ground state eigenenergies are presented in Appendix B. Hence we focus only on 
the ground state energies ($\mathcal{E}^0\sim E^0$) for the rest of the analysis.

\begin{figure}[h]
\includegraphics[height=3.5cm, width=8.5cm]{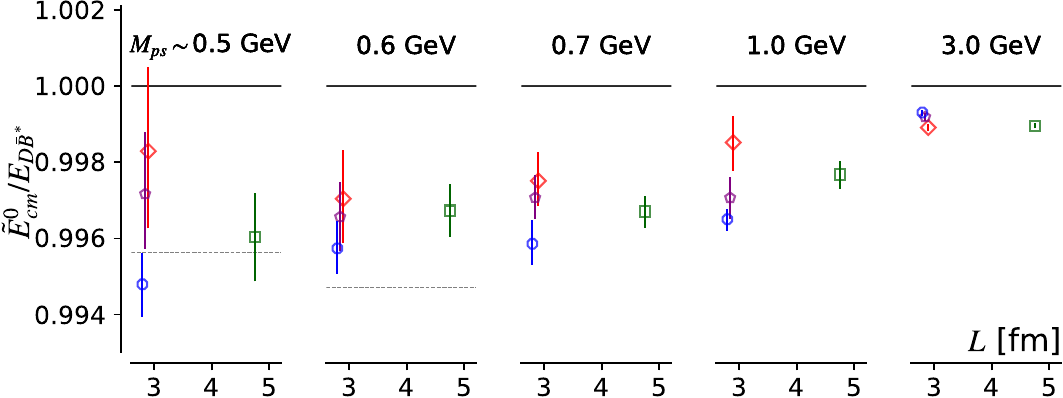}
\caption{The ground state energies in units of $E_{D\bar B^*}$ on all ensembles (see \tbn{lattice} 
for color-symbol conventions) for all $M_{ps}$ values (different vertical panels).}
\eef{gsspectrum}

In \fgn{gsspectrum}, we present the ground state energy estimates, in units of $E_{D\bar B^*}$, 
at various $M_{ps}$ and for all the ensembles. These estimates are evaluated as 
$E^n = \Delta E^n + M_{D} + \tilde{M}_{\bar{B}^*}$, where $\Delta E^n$ is the estimate from fit 
to $R^n(t)$, $\tilde{M}_{\bar B^{*}} = M_{\bar B^{*}}- 0.5\overline M^{\bar bb}_{lat} $ 
$+ 0.5 \overline M^{\bar bb}_{phys}$ accounts for the NRQCD additive offset, and 
$\overline M^{\bar bb}_{lat}(\overline M^{\bar bb}_{phys})$ refers to the spin averaged mass of 
the $1S$ bottomonium measured on the lattice (experiments). The eigenenergies clearly show 
a trend of decreasing energy spitting, hence decreasing interaction strength, with increasing 
$M_{ps}$. Another interesting feature to note here is the nonzero lattice spacing ($a$) 
dependence of the ground state energies on similar volume ensembles ($S_1$, $S_2$, $S_3$), 
which we account for through an $a$ dependence in the parametrized amplitude as discussed below. 

In \fgn{gsspectrum}, we also indicate the branch point location of the left hand cut (lhc) arising 
out of an off-shell pion exchange process for different $M_{ps}$ by horizontal dashed lines.
Recent developments point to the importance of lhc effects on virtual subthreshold poles related 
to the $T_{cc}$ tetraquark \cite{Du:2023hlu}. Such effects on bound states are the subject of future 
studies where one could successfully solve the relevant three particle integral equations. This is 
beyond the scope of this work and we ignore such effects in our analysis.

{\it $D\bar B^*$ scattering amplitude}: Assuming these energy splittings in ground states are purely 
described by elastic scattering in the $D\bar B^*$ system, we utilize them to constrain the associated 
$S$-wave scattering amplitude following L\"uscher's finite-volume prescription \cite{Luscher:1990ux,
Briceno:2014oea}. For the low energy scattering of $D$ and $\bar B^*$ mesons, where other multi-particle 
thresholds are sufficiently high \cite{Draper:2021clv}, in the $S$-wave leading to the total angular 
momentum and parity $J^P=1^+$, the scattering phase shifts $\delta_{l=0}(k)$ are related to the 
finite-volume energy spectrum through $kcot[\delta_0(k)] = 2Z_{00}[1;(\frac{kL}{2\pi})^2)]/(L\sqrt{\pi})$. 
Here, $k$($E_{cm}=\sqrt{s}$) is the momentum (energy) in the center of momentum frame such that 
$4sk^2 = (s-(M_{D}+M_{\bar B^*})^2)(s-(M_{D}-M_{\bar B^*})^2)$. We follow the procedure outlined in 
Appendix B of Ref. \cite{Padmanath:2022cvl} to constrain the amplitude. A sub-threshold pole in 
the $S$-wave scattering amplitude $t = ({\mathrm{cot}}\delta_0 - i)^{-1}$ occurs when 
$k{\mathrm{cot}}\delta_0 = \pm\sqrt{-k^2}$ for scattering in $S$-wave. 

We parametrize the elastic $D\bar B^*$ scattering amplitude in terms of the scattering length $a_0$ 
in an effective range expansion near the threshold, supplemented by a lattice spacing dependence. 
This is required to incorporate the cut-off effects observed in the ground state energy estimates. 
We find that a linear functional form given by $k{\mathrm{cot}}\delta_0 = A^{[0]} + aA^{[1]}$, where 
$A^{[0]}=-1/a_0$, accommodate the $a$ dependence of the $k{\mathrm{cot}}\delta_0$ estimates. We 
present the fit results for $A^{[0]}=-1/a_0$ in \fgn{a0a1_separate} (circle symbols) as a function 
of $M_{ps}$ involved.  Alternative fitting choices with a leading quadratic dependence or using 
only data from non-charm $M_{ps}$ are consistent with results in \fgn{a0a1_separate}, see details 
in Supplemental material \cite{Suppl}. 

The sign of $A^{[0]}=-1/a_0$ determines the fate of the near-threshold pole, if there exists one. 
A negative (positive) value of $A^{[0]}$($a_0$) indicates that the interaction potential is strong 
enough to form a real bound state\cite{Landau:1991wop}. After considering all possible systematics, 
we find that for all the non-charm light quark masses, $A^{[0]}$ is negative, which indicates an 
attractive interaction strong enough to host a real bound state. On the contrary, at the charm point, 
despite the unambiguous negative energy shifts in the ground states, the attraction is weak to host 
any real bound state as suggested by the positive value of $k{\mathrm{cot}}\delta_0$ in the continuum 
limit. This observation goes in line with the phenomenological expectation for doubly heavy four 
quark ($QQ'l_1l_2$) systems with $m_{l_1}=m_{l_2}$ that the binding increases with increased relative 
heaviness of the heavy quarks with respect to its light quark content\cite{Francis:2016hui,Czarnecki:2017vco,Junnarkar:2017sey}. 

Now we investigate the light quark mass ($m_{u/d}$) or $M_{ps}$ dependence of the fitted parameters. 
To this end, we consider three different parametrizations: a linear dependence ($f_l(M_{ps}) = 
\alpha_c + \alpha_l M_{ps}$) to probe the heavy light quark mass case, a leading $M_{ps}^2$ 
dependence ($f_s(M_{ps}) = \beta_c + \beta_s M_{ps}^2$) to assess the chiral behaviour, and a 
quadratic dependence ($f_q(M_{ps}) = \theta_c + \theta_l M_{ps} + \theta_s M_{ps}^2$) to quantify 
the associated systematics. In \fgn{a0a1_separate}, we show the fit results for this $M_{ps}$ 
dependence in colored bands. The two stars represent $A^{[0]}$ at the physical $M_{ps}$ 
(equivalently the physical scattering length $a_0^{phys}$) and the critical $M_{ps}$ at which 
$A^{[0]}$ changes its sign or above which the system becomes unbound. It is indeed desired to have 
more points in the intermediate mass regime between the charm and the strange 
\begin{figure}[h]
\includegraphics[scale=0.46]{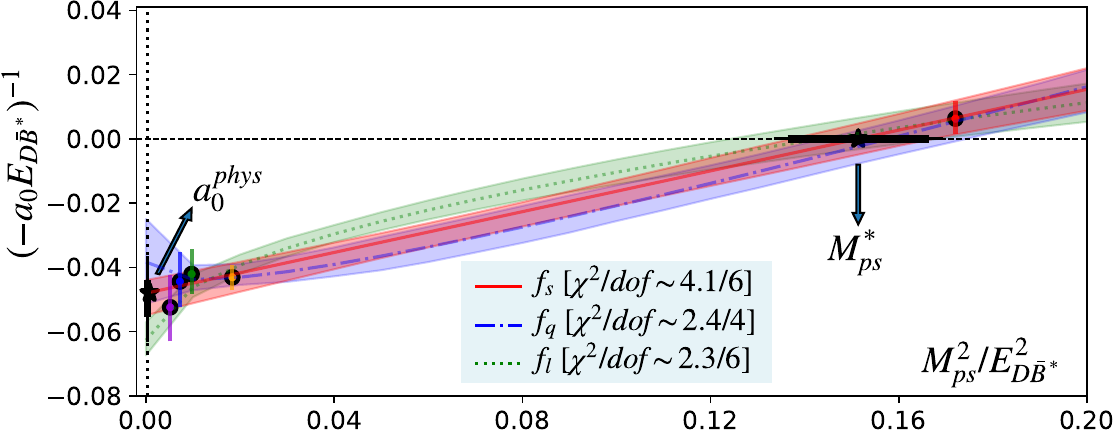}
\caption{Continuum extrapolated $k{\mathrm{cot}}\delta_0$ or $A^{[0]}=-1/a_0$ estimates of the $D\bar B^*$ system 
as a function of $M_{ps}^2$ in units of $E_{D\bar B^*}$. The dotted vertical line close to the $y$-axis indicates 
$M_{ps}=M_{\pi}^{phys}$. The two star symbols represent the amplitude at $M_{ps} = M_{\pi}^{phys}$ and 
the critical $M_{ps}=M^{*}_{ps}$ above which the system becomes unbound.}
\eef{a0a1_separate}
quark masses to further constrain the dependence. Yet, our fits demonstrate near independence in 
the fit forms as can be observed from the consistency between the error bands from different fit 
forms. 

Based on the fit form $f_s(M_{ps})$ in the chiral regime, we find that the scattering length of the $D\bar B^*$ 
system at the physical light quark mass ($m_{u/d}^{phys}$), corresponding to $M_{ps}=M_{\pi}^{phys}$, to be
\beq
a_0^{phys} = 0.57(^{+4}_{-5})(17) \mbox{~fm}.
\eeq{scatlen}
The asymmetric errors indicate the statistical uncertainties, whereas the second parenthesis quotes 
the systematic uncertainties with the most dominant contribution arising from the chiral extrapolation 
fit forms. The positive value of the scattering length at $M_{ps}=M_{\pi}^{phys}$, at the level of
3$\sigma$ uncertainty, is an unambiguous evidence for the strength of the $D\bar B^*$ interaction 
potential to host a real $bc\bar u\bar d$ tetraquark bound state $T_{bc}$ with binding energy 
\beq
\delta m_{T_{bc}} = -43(^{+6}_{-7})(^{+14}_{-24}) \mbox{~MeV},
\eeq{betbc}
with respect to $E_{D\bar B^*}$. The first parenthesis indicates the statistical errors and the second one 
quantifies various systematic uncertainties added in quadrature. The pseudoscalar meson mass, corresponding 
to the critical light quark mass, where $a_0$ diverges, is found to be $M^{*}_{ps} = 2.73(21)(19) \mbox{~GeV}$. 
This critical point also signifies that QCD dynamics within such exotic systems is such that at a heavy 
light quark mass the system of quarks perhaps reaches the unitary gas limit, as indicated by the 
divergent scattering length \cite{Newton:1982qc}. For $M_{ps}\ge M^{*}_{ps}$, the $T_{bc}$ system 
remains unbound.

{\it Systematic uncertainties}: Our lattice setup together with the bare bottom and charm quark mass 
tuning procedure has been demonstrated to reproduce the $1S$ hyperfine splittings in quarkonia with 
uncertainties less than 6 MeV \cite{Mathur:2022ovu,Mathur:2016hsm}. We observe the effects of such a 
mistuning of either of the heavy quark mass on the energy splittings we extract are very small compared
to the statistical errors. Additionally, our strategy of evaluating the energy differences and working 
with mass ratios has also been shown to significantly mitigate the systematic uncertainties related to 
heavy quark masses \cite{Mathur:2018epb,Mathur:2022ovu}. This is observed to be the case in this study 
as well, leading to transparent signals for the ground state energy as shown in Figures \ref{fg:fitcompare},
\ref{fg:spectrum}, and \ref{fg:gsspectrum}. Our fitting procedure involves careful and conservative 
determination of statistical errors, and uncertainties related to the excited-state-contamination and 
fit-window errors. Additional checks using alternative quark smearing procedures also agree with our 
energy estimates, see Appendix A. The amplitude determination and followed extrapolations are performed 
with results from varying the fit-windows to evaluate the uncertainties propagated to our final results. 
The uncertainties related to the fit forms used in chiral extrapolations are observed to be the most 
dominant, as is evident from Figure \ref{fg:a0a1_separate}. We assume the partially quenched setup
involving ensembles with different sea pion masses, we utilize, have negligible effects on the energy 
splittings we extract for the explicitly exotic $T_{bc}$ tetraquark, similar to what was observed for 
heavy hadrons in Refs. \cite{Dowdall:2012ab,McNeile:2012qf}. Uncertainty related to scale setting is 
also found to be negligible in comparison to the statistical uncertainties in the energy splittings.

{\it Summary}: We have performed a lattice QCD simulation of coupled $D\bar B^*$-$\bar BD^*$ scattering with 
explicitly exotic flavor $bc\bar u\bar d$ and $I(J^P) = 0(1^+)$. Following a rigorous extraction of 
finite-volume eigenenergies and continuum extrapolated elastic $D\bar B^*$ scattering amplitudes for 
the five light quark masses studied, we determine the light quark mass dependence of the elastic 
$D\bar B^*$ scattering length $a_0$. We observe unambiguous negative energy shifts between the interacting 
and non-interacting finite-volume energy levels. Our estimate for $a_0^{phys}$ (\eqn{scatlen}) is 
positive, indicating an attractive interaction between the $D$ and $\bar B^*$ mesons, which is strong 
enough to host a real bound state with binding energy $\delta m_{T_{bc}} = -43(^{+6}_{-7})(^{+14}_{-24})$ 
MeV. We find that the strength of interaction is such that this $bc\bar u\bar d$ tetraquark becomes 
unbound at $M^{*}_{ps}$, which is close to the $\eta_c$ meson mass. 

In this work, we make several important steps ahead to arrive at robust inference on the nature of interaction
between the $D$ and $\bar B^*$ mesons. Our main strategy has been to determine the signature of scattering length 
in $D\bar B^*$ interactions at the physical pion mass $a_0^{phys}$. Our results indicate that $a_0^{phys}$ is 
positive, which suggests that attractive $D\bar B^*$ interactions are strong enough to host a real bound state. 
Further theoretical investigations are desired to reduce the uncertainties in the binding energy of $T_{bc}$ 
with respect to $E_{D\bar B^*}$. Fully dynamical simulations on several more ensembles, with different volumes 
and improvized fermion actions, high statistics studies with lighter $m_{u/d}$, etc. are a few other 
improvisations that can further constrain the relevant scattering amplitude. Additionally, future works 
involving Hermitian correlation matrices at rest as well as in moving frames and those using bilocal two-meson 
interpolators with nonzero relative meson momenta aimed at reliable excited state extraction would be 
a few important steps ahead \cite{Padmanath:2018tuc,Padmanath:2022cvl,Chen:2022vpo,Wagner:2022bff}. We 
hope that our observations and inferences in this work will motivate more theoretical efforts and 
experimental searches for such states.

\begin{acknowledgments}
This work is supported by the Department of Atomic Energy, Government of India, under Project Identification Number RTI 4002. M.P. gratefully acknowledges support from the Department of Science and Technology, India, SERB Start-up Research Grant No. SRG/2023/001235. We are thankful to the MILC collaboration and in particular to S. Gottlieb for providing us with the HISQ lattice ensembles. We thank Sara Collins for a careful reading of the manuscript. We thank the authors of Ref. \cite{Morningstar:2017spu} for making the {\it TwoHadronsInBox} package utilized in this work. We also thank Gunnar Bali, Parikshit Junnarkar, Alexey Nefediev, Sayantan Sharma, Stephen R. Sharpe, and Tanishk Shrimal for discussions. Computations were carried out on the Cray-XC30 of ILGTI, TIFR. Amplitude analyses were performed on Nandadevi computing cluster at IMSc Chennai. N. M. would also like to thank A. Salve and K. Ghadiali for computational support.
\end{acknowledgments}


\textit{Appendix A: Ground state energy plateau.-}
In this work we have utilized a wall-source point-sink setup to construct the necessary two-point 
correlation functions. The use of such an asymmetric setup implies the effective energies 
$aE_{eff} = [ln(C(t)/C(t+\delta t))]/\delta t$ could approach their asymptotic values as rising-from-below, 
due to the nonpositive definite nature of the coefficients in a spectral decomposition, in 
contrast to a falling-from-above behaviour in a symmetric setup. In \fgn{boxsnk_plot}, we show 
the effective mass in wall-source point-sink setup with the brown-circle ($R^2 = 0$) which rises from below.

To avoid any ambiguity in selecting the plateau regions of effective masses of such correlators, we also 
employ a wall-source box-sink setup \cite{Hudspith:2020tdf}, which asymptotically approaches the symmetric 
limit. In the symmetric limit, the effective masses are expected to follow a conventional falling-from-above 
feature, modulo the statistical noise. To this end, we vary the smearing radius $R$ to investigate 
the time dependence of effective mass plateaus in the approach to the symmetric limit. In \fgn{boxsnk_plot}, 
we present a comparison of the effective energy (top) and effective energy splittings (bottom) determined 
using different quark sink smearing procedures for the case of $M_{ps}\sim700$ MeV on the finest ensemble. 
Clearly the rising-from-below behaviour is gradually disappearing in the approach to the symmetric limit. 
It is also evident that the results at the large time limit from point-sink and box-sink are very much 
consistent with each other affirming our assessment on effective mass plateau in choosing a fit range. Such 
a behavior of effective masses with varying smearing radii was also observed in Ref.~\cite{Hudspith:2020tdf}. 
In the large time limit, where the signal quality is still good, all of sink smearing cases suggest 
consistent negative energy shifts. This is evident from the large time behaviour of energy splittings 
presented in the bottom panel of \fgn{boxsnk_plot}, where the correlated statistical noise, not related to 
the excited state contamination, is suppressed between the numerator and denominator in the ratio correlators $R^n(t)$.

The agreement of energy splitting estimates from fits to $R^n(t)$ with those evaluated from separate fits 
to the GEVP eigenvalue correlators $\lambda^n(t)$ and the single-meson correlators $\mathcal{C}_{D/\bar B^*}$ 
at large times (see \fgn{fitcompare}) already rules out the usual concern of accidental partial cancellation 
of excited state contaminations in $R^n(t)$. The consistency at large times between ground state energy 
plateaus from different sink-smearing radii observed in top panel of \fgn{boxsnk_plot} further affirms 
the reliable isolation of the ground state plateau. Note also that the magnitude of such cancellations and 
the ground state saturation times could be different in different lattice QCD ensembles. All the ground 
state estimates for noncharm $M_{ps}$ values in our study are determined from the time intervals approximately 
between 1.5(2) fm [$t_{min}$] to 2.3(2) fm [$t_{max}$]. The consistent ground state saturation times across 
different ensembles with different specifications further imply the reliability of our ground state saturation, 
despite our asymmetric setup.

\bef[hbt]
\includegraphics[scale=0.45]{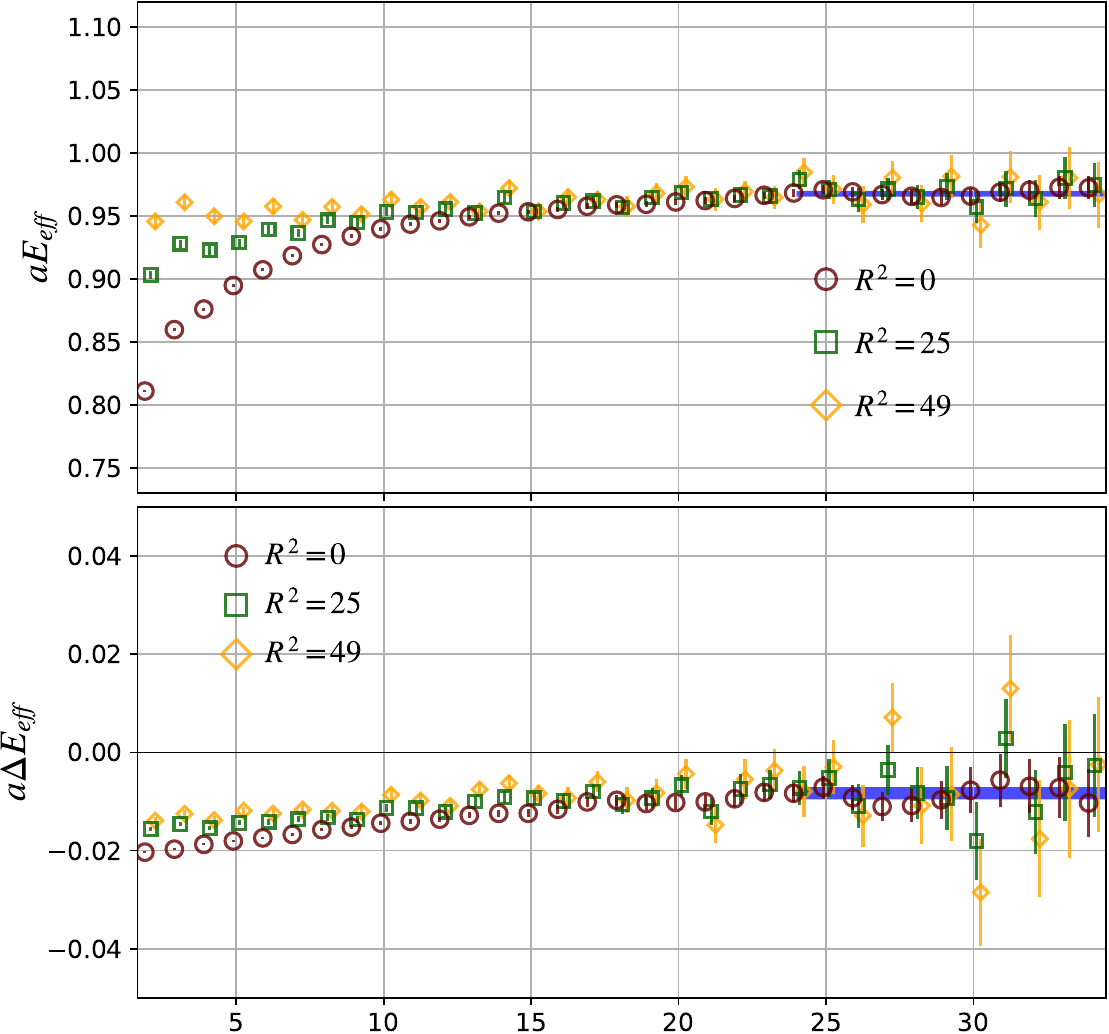}
\caption{Comparison of effective energy (top) and effective energy splitting (bottom) for the ground state as 
determined using three different smearing radii applied on the quark fields at the sink timeslice. The legend 
indicates the smearing radius squared in units of the lattice spacing \cite{Hudspith:2020tdf}. The blue horizontal 
band indicates the final fit estimate for the energy and energy splitting. The results presented are for 
the case $M_{ps}\sim700$ MeV on the finest ensemble. }
\eef{boxsnk_plot}

\textit{Appendix B: Elastic $D\bar B^*$ excitations.-} Gaining access to higher lying elastic excitations 
in the $D\bar B^*$ channel is an important step ahead towards constraining the energy dependence of the 
amplitude over a long energy range. However, within the wall-smearing setup, all the nonzero momentum 
excitations are significantly suppressed. This suppression is exact in a free theory, and is empirically 
confirmed from the early plateauing and from the quality of signals in the interacting theory. While this 
suppression is advantageous in ground state energy determination (see Refs.\cite{Francis:2016hui,Francis:2018jyb,
Mathur:2018epb,Junnarkar:2018twb,Hudspith:2020tdf,Mathur:2022ovu} for details), the suppressed coupling to 
the nonzero momentum excitations implies that the access to higher two-meson elastic excitations with 
nonzero relative meson momenta are restricted in the wall-smearing setup. This implies other methodologies 
that facilitate the use of bilocal two-meson interpolators with separately momentum projected mesons are 
necessary in future studies \cite{HadronSpectrum:2009krc,Abdel-Rehim:2017dok,Wagner:2022bff}\footnote{While this 
letter was being reviewed, a preprint, Ref.~\cite{Alexandrou:2023cqg} appeared which utilizes bilocal 
two-meson interpolators in their analysis, utilizing the methods in Ref.~\cite{Abdel-Rehim:2017dok}.}. In 
this respect, it is informative to know the location of the lowest non-interacting level with nonzero relative 
meson momenta and whether it is close enough to influence the ground state energies in any substantial way. 
Considering this, in \fgn{gsspectrumee} we present the ground state eigenenergies along with the $D\bar B^*$ 
threshold and the next lowest elastic $D\bar B^*$ excitation with nonzero relative meson momentum determined 
using the continuum dispersion relation that is assumed in the finite-volume quantization condition 
\cite{Luscher:1990ux,Briceno:2014oea}. Clearly, the location of this first non-interacting elastic excitation 
is sufficiently high to have any nonnegligible effects on the extracted the ground state energies.

\begin{figure}[h]
\includegraphics[height=3.5cm, width=8.5cm]{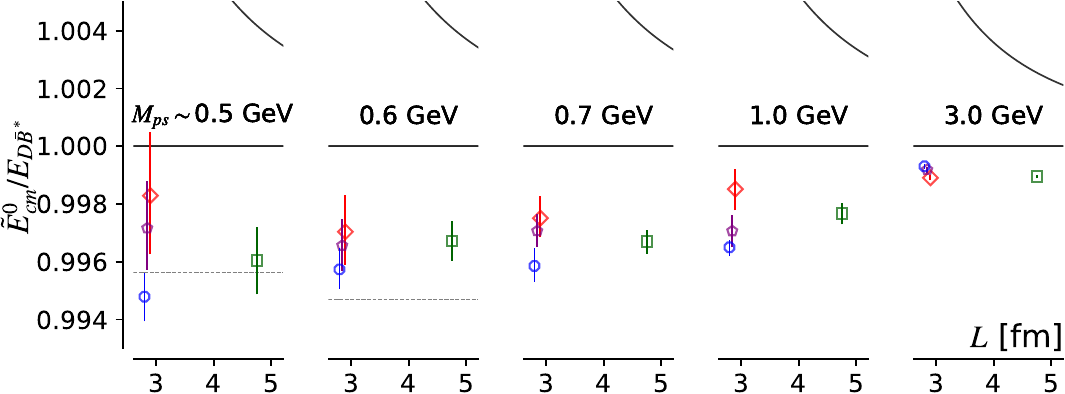}
\caption{The ground state energy eigenvalues in the background of lowest two non-interacting $D\bar B^*$ 
finite-volume levels units of $E_{D\bar B^*}$ on all ensembles (see \tbn{lattice} for color-symbol 
conventions) for all $M_{ps}$ values (different vertical panels).}
\eef{gsspectrumee}

\bibliography{paper}


\onecolumngrid
\clearpage
\onecolumngrid

\begin{center}
  {\Large \bf Supplemental material}
\end{center}

\makeatletter
\c@secnumdepth=4
\makeatother

\newif\ifsepsupp
\sepsuppfalse

\input{supplement.tex}

\end{document}

%% file: supplement.tex
\setcounter{equation}{0}
\setcounter{figure}{0}
\setcounter{table}{0}

\section{Low lying eigenenergies on all ensembles}\label{spectrum}
\begin{figure}[thb]
\includegraphics[scale=0.261]{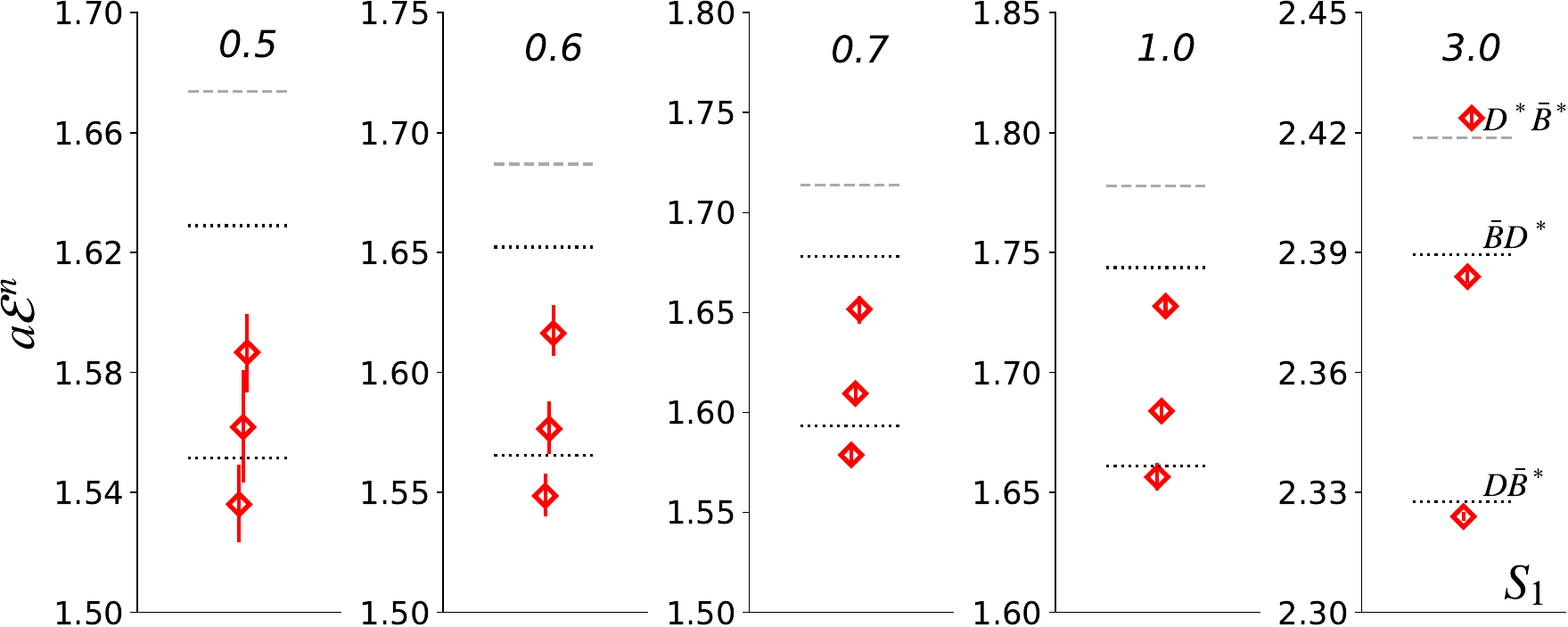}\hspace{0.2cm}
\includegraphics[scale=0.261]{./spectrum_l40.pdf}\\\vspace{0.3cm}
\includegraphics[scale=0.261]{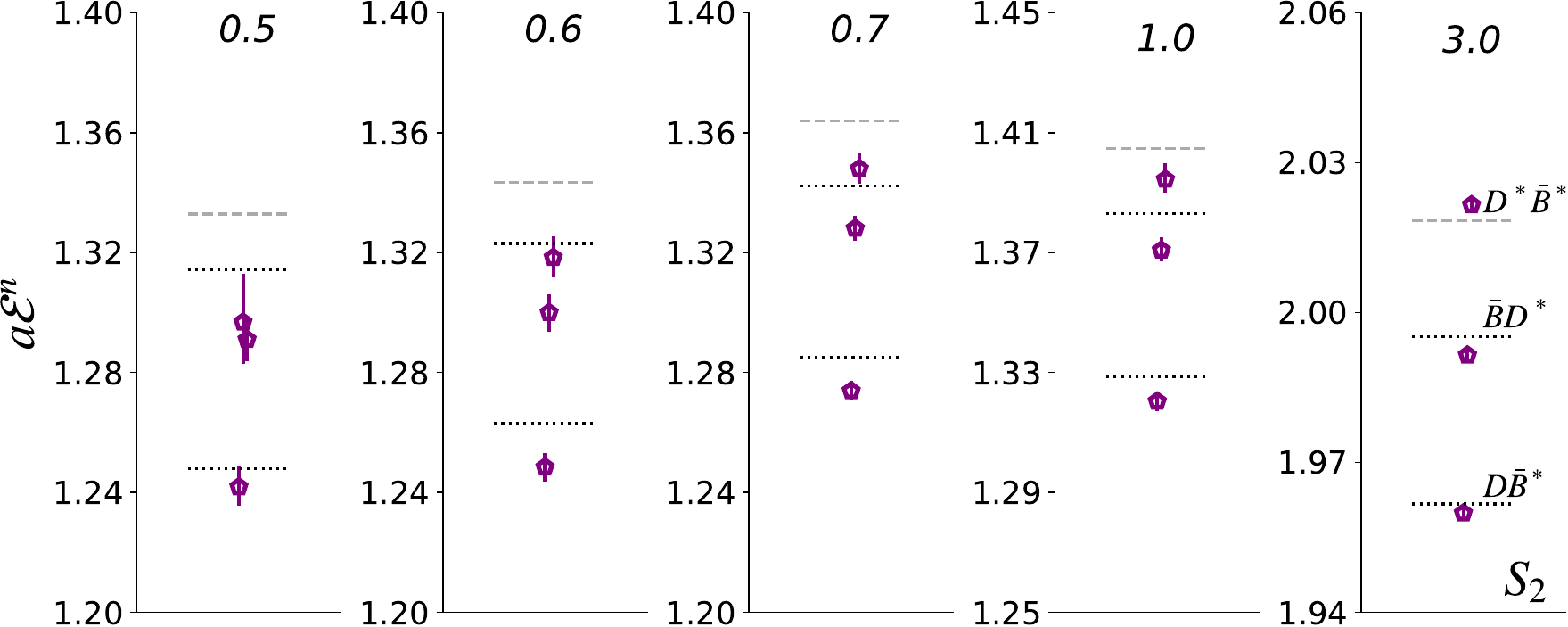}\hspace{0.2cm}
\includegraphics[scale=0.261]{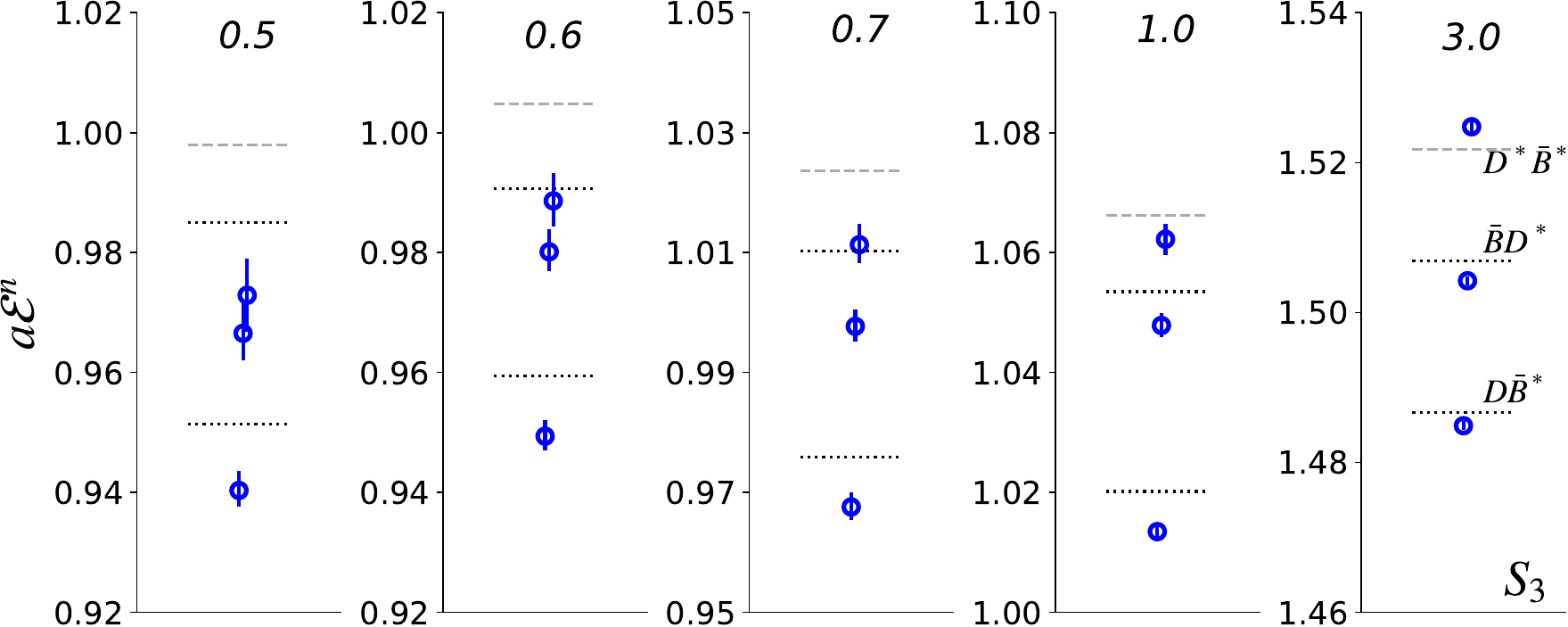}
\caption{Low-lying finite-volume eigenenergies for isoscalar axialvector $bc\bar u\bar d$
channel on the four ensembles studied. The near degenerate eigenlevels are slightly 
shifted horizontally for clarity.}
\label{fg:spectrum}
\end{figure}
In this section, we present the finite-volume GEVP eigenenergies of the isoscalar axialvector 
$bc{\bar{u}}{\bar{d}}$ channel that we extract on all four ensembles listed in Table I of 
the main article, at the five different $m_{u/d}$ values corresponding to $M_{ps}\sim$ 
0.5, 0.6, 0.7, 1.0, and 3.0 GeV. The eigenenergies shown in lattice units include 
the additive offsets related to the NRQCD-based dynamics of heavy bottom quarks. 
The non-interacting two-meson energy levels corresponding to $D\bar B^*$ and $\bar BD^*$ thresholds 
are indicated as dotted horizontal line segments for each lattice and each $M_{ps}$. 
The $D^*\bar B^*$ threshold in each case is also shown in the figure by dashed lines.
Note that the use of wall-smearing setup restricts any direct access to the elastic two-meson 
excitations with nonzero relative meson momenta. This means although a reliable ground state 
extraction could be made, the excited eigenenergies may not represent the real elastic excitations.

\section{Operator-state overlaps}\label{sec:OSO}
\bef[htb]
\includegraphics[scale=0.65]{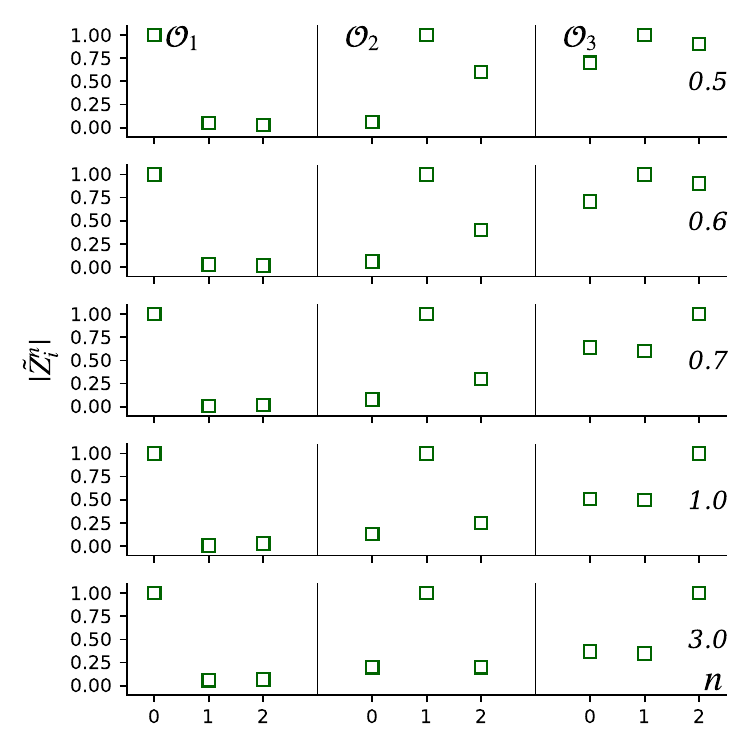}
\caption{Modulus of normalized sink operator-state overlaps $|\tilde{Z}_i^n|$ for an eigenenergy 
indicated by $n={0, 1, 2}$ and an operator represented by $\mathcal{O}_i$, where $i={1, 2, 3}$ on 
the $L_1$ ensemble. The errors in the normalized overlap factors are smaller than the symbols 
and hence are suppressed. }
\eef{Zratiosl40}
In \fgn{Zratiosl40}, we present the modulus of normalized sink operator-state overlaps $|\tilde{Z}_i^n|$, 
normalized such that its largest value for any given operator $\mathcal{O}_i$ across all the eigenenergies
$\{n\}$ is unity \cite{Dudek:2009qf,Padmanath:2013zfa}. $\tilde{Z}_i^n$ quantifies the relative 
relevance of any given operator across all the eigenenergies. The $|\tilde{Z}_i^n|$ values are presented for 
all $M_{ps}$ cases on the $L_1$ ensemble. Each square marker corresponds to the $|\tilde{Z}_i^n|$ for 
a given operator $\mathcal{O}_i$ on to a given eigenenergy $n$. Each horizontal panel stands for an $M_{ps}$ 
indicated on the right-hand side, whereas the vertical lines in each horizontal panel part $|\tilde{Z}_i^n|$
for different operators indicated on the top panel. The $x$-axis ticks refer to the three finite-volume 
eigenenergies we have extracted. $\mathcal{O}_1$, the two-meson operator related to $D\bar B^*$ threshold, 
can be seen to have the largest overlap with the ground state and has significantly small overlaps with 
the excited eigenenergies. $\mathcal{O}_2$, the two-meson operator related to $\bar BD^*$ threshold, has the largest 
overlap with the first excited eigenenergy and a very small overlap with the ground state. $\mathcal{O}_2$ 
also have nonnegligible overlap factors with the second excited eigenenergy indicating $\bar BD^*$-type 
two-meson Fock component, which decreases with increasing $M_{ps}$. On the other hand, $\mathcal{O}_3$, 
the diquark-antidiquark type operator, have substantial overlap factors with all eigenenergies at the two 
lightest $M_{ps}$ values, whereas with an increased $M_{ps}$ its largest overlap is with the second excited 
eigenenergy. Note that $\mathcal{O}_3$ is Fierz related to two-meson interpolators \cite{Padmanath:2015era}, 
and the large $\tilde{Z}_3^n$ values of $\mathcal{O}_3$ for all $n$ could be related to this underlying 
connection between two-meson and diquark-antidiquark operators.

A summary from the above observations on overlap factors is as follows. $\mathcal{O}_1$ predominantly 
determines the ground state, whereas it has significantly small coupling with the excited eigenenergies. 
Similar patterns of overlap factors are also observed for other ensembles, all of which indicate 
that $\mathcal{O}_1$ predominantly determines the ground state. The two excited eigenenergies have strong 
two-meson and diquark-antidiquark Fock components in the two lightest $M_{ps}$ values. One could also 
evaluate and investigate the normalized source operator-state overlaps from the left eigenvectors of 
$\mathcal{C}$ in Eq. (1) in the main draft, which also leads to the same conclusions.

\section{Operator basis dependence}
\bef[tbh!]
\includegraphics[scale=0.55]{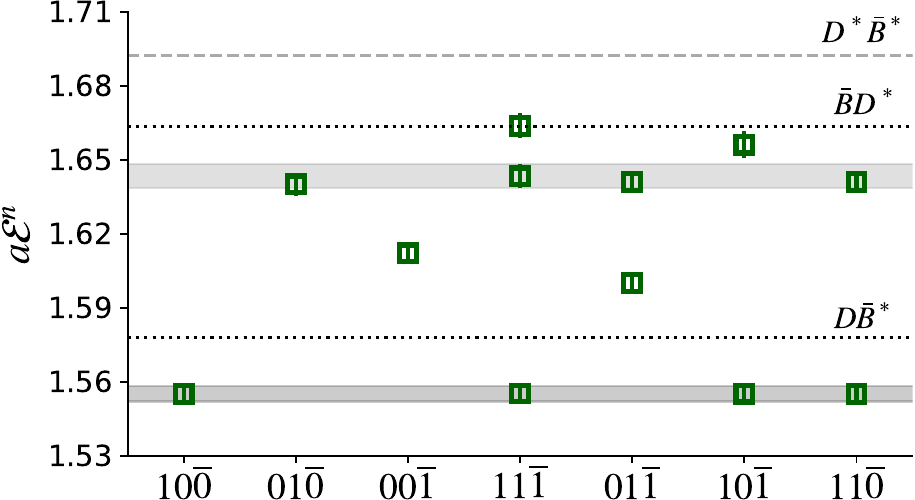}
\caption{Operator basis dependence of the low lying eigenenergies of the $L_1$ ensemble and
$M_{ps}\sim$700 MeV for all possible operator basis that can be built out of the three operators
utilized in this work. }
\eef{basisdep}
In \fgn{basisdep}, we show the operator basis dependence as determined for $M_{ps}\sim$
700 MeV in the $L_1$ ensemble, for various operator basis build out of $\mathcal{O}_1$,
$\mathcal{O}_2$, and $\mathcal{O}_3$ operators as defined in Eq. (2) of the main draft. 
The digital indexing on the $x$-axis tick labels refers to various operator basis in the 
order $\{\mathcal{O}_1, \mathcal{O}_2, \mathcal{O}_3\}$, with an overline on the third 
index as a visual aid within the plot to highlight the diquark-antidiqaurk interpolator. 
1 (0) indicates an operator is included in (excluded from) the basis. The horizontal
lines refer to the $D\bar B^*$, $\bar BD^*$ and $\bar B^*D^*$ thresholds. The gray horizontal bands
refer to the two lowest levels in the full basis indicated by $11\overline{1}$. A level
below the threshold appears only when $\mathcal{O}_1$ is present in the basis. The first
excited eigenenergy in the full basis $11\overline{1}$ is faithfully reproduced in those bases
where $\mathcal{O}_2$ is included. $\mathcal{O}_3$ alone does not precisely determine
any level among the GEVP eigenenergies using full basis. Similar observations are also made
on other ensembles. In summary, the ground state in the full basis $11\overline{1}$ is 
reliably determined with $\mathcal{O}_1$ and is unaffected by the inclusion of $\mathcal{O}_2$ 
and $\mathcal{O}_3$ operators. 

\bet[hbt!]
  \begin{center}
	  \begin{tabular}{p{2.0cm}p{2.0cm}p{1.5cm}>{\hfill\arraybackslash}p{2.cm}>{\hfill\arraybackslash}p{2.cm}>{\hfill\arraybackslash}p{2.5cm}>{\hfill\arraybackslash}p{2.cm}}
      \hline
		  $M_{ps}$ [GeV] & $\chi^2/d.o.f$ & \multicolumn{2}{c}{Linear} & $\chi^2/d.o.f$ & \multicolumn{2}{c}{Quadratic}\\
      
		  & & $A^{[0]}/E_{D\bar B^*}$ & $A^{[1]}/E_{D\bar B^*}$ && $A^{[0]}/E_{D\bar B^*}$ & $A^{[2]}/E_{D\bar B^*}$\\\hline
		  0.5 & 2.1/2 & $-0.05(1)$ & $~0.17(_{-11}^{+13})$ & 2.1/2 & $-0.045(6)$ & $0.9(_{-6}^{+7})$  \\\hline
		  0.6 & 0.5/2 & $-0.044(_{-8}^{+9})$ & $~0.10(_{-9}^{+9})$ & 0.5/2 & $-0.040(_{-5}^{+6})$ & $0.6(5)$ \\ \hline
		  0.7 & 3.0/2 & $-0.042(_{-6}^{+8})$ & $~0.09(_{-7}^{+6})$ & 3.7/2 & $-0.037(_{-4}^{+5})$ & $0.5(_{-4}^{+3})$ \\ \hline
		  1.0 & 2.9/2 & $-0.043(4)$ & $~0.11(_{-5}^{+5})$ & 2.9/2 & $-0.038(_{-3}^{+3})$  & $0.8(3)$  \\ \hline
		  3.0 & 3.6/2 & $~0.006(_{-5}^{+6})$ & $-0.20(_{-5}^{+4})$ & 3.6/2 & $-0.002(_{-3}^{+4})$ & $-1.2(_{-3}^{+2})$ \\ \hline
      \hline
  \end{tabular}
  \end{center}
\caption{Results from amplitude fits at different light quark mass cases indicated
in terms of $M_{ps}$ in the first column. The amplitude is approximated to be determined 
by the scattering length, with a linear or quadratic lattice spacing dependence as discussed 
in the main draft. The optimized parameter values in the table are presented in units of 
energy of the $D\bar B^*$ threshold, $E_{D\bar B^*}$. The $A^{[0]}$ parameter in either parametrization 
is negative of the inverse scattering length in the continuum limit. }
\eet{Ampfits1}

\begin{figure}[hbt!]
\includegraphics[scale=0.5]{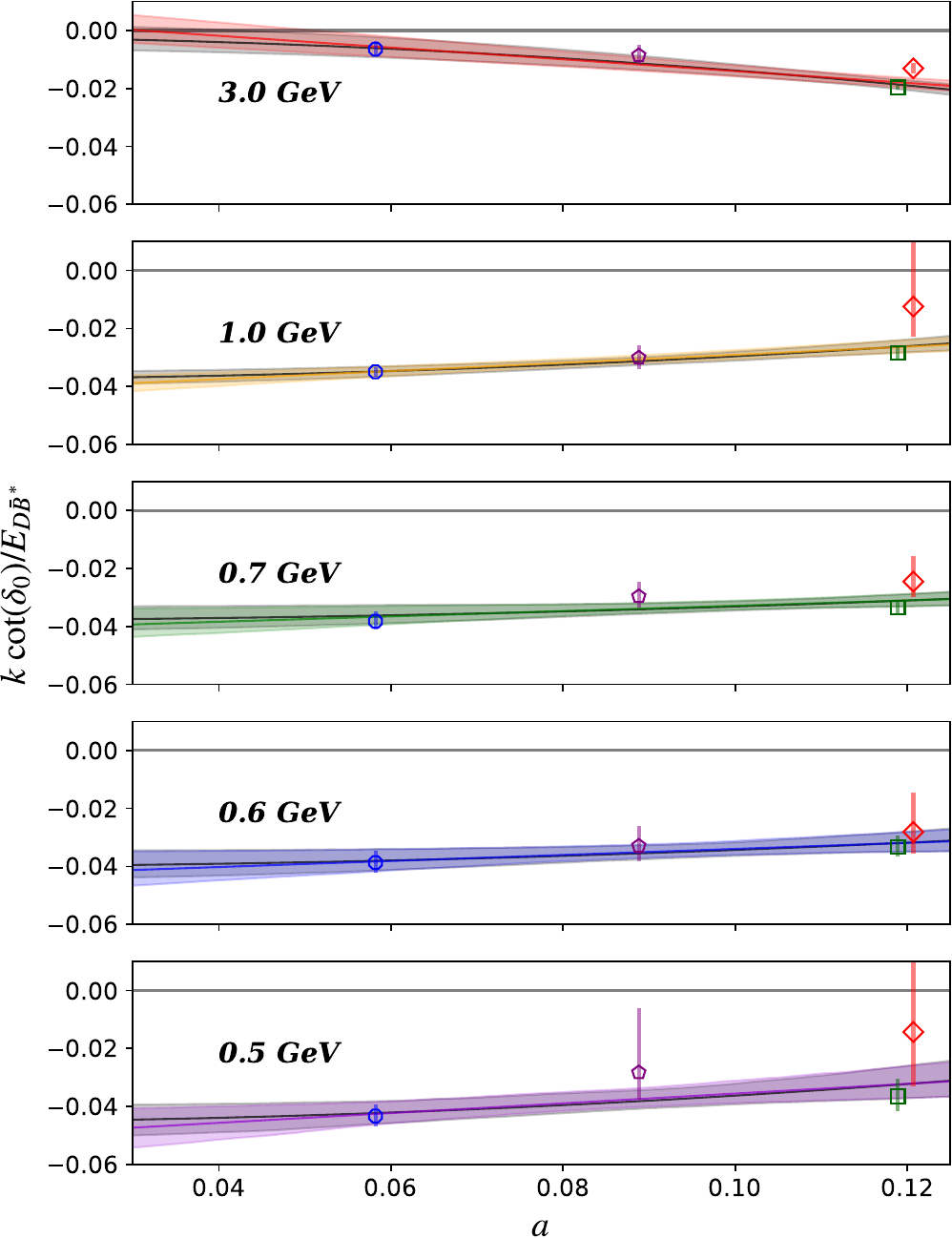}
\includegraphics[scale=0.5]{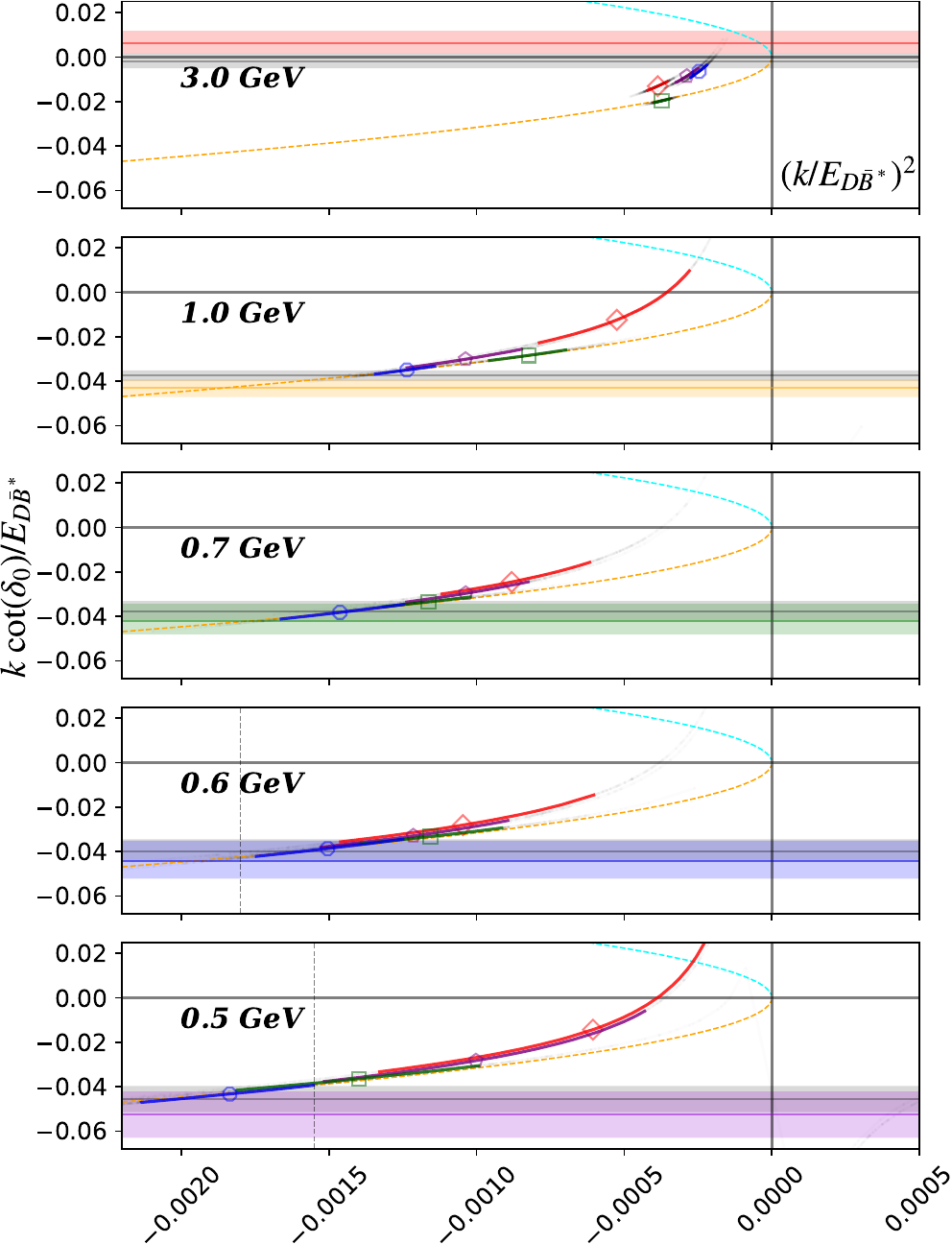}
\caption{Left: $k{\mathrm{cot}}\delta_0$, in units of the elastic threshold $E_{D\bar B^*}$, versus $a$
(lattice spacing) for all $M_{ps}$ values. We follow the marker/color coding in Table I of 
the main draft for the data points referring to the simulated data. The colored/gray bands 
indicate the fit results with linear and quadratic lattice spacing dependence, respectively. 
Right: $k{\mathrm{cot}}\delta_0$ versus $k^2$ for all $M_{ps}$ values studied in units of 
the elastic threshold $E_{D\bar B^*}$. The dashed orange (cyan) curve indicates the constraint for 
the existence of a sub-threshold pole in the scattering amplitude. The horizontal bands are 
the continuum extrapolated estimates of $k{\mathrm{cot}}\delta_0$ for the respective $M_{ps}$. 
The black dashed vertical lines in the plots on the right indicate the location of the branch
point associated with the left hand cut arising from the $D\bar B\pi$ channel. }
\eef{pcotdelta_summary}

\section{Results on scattering amplitude}
In \tbn{Ampfits1}, we tabulate the results from different amplitude fits that were performed.
In \fgn{pcotdelta_summary}, we present the quality of these fits by comparing the fit results 
with the data points (see the figure caption for details). On the left of \fgn{a0_mpi2_fits_LQ}, 
we present the light quark mass dependence in the chiral regime determined from continuum 
extrapolated elastic $D\bar B^*$ scattering amplitudes following a linear and quadratic lattice 
spacing dependence. We present the final estimate (black star) from the linear fit form 
considering the presence of heavy quarks in our system, whereas the difference in the fit 
results are accounted in the systematics quoted in the main draft. On the right of \fgn{a0_mpi2_fits_LQ}, 
we present a comparison of the light quark mass dependence in the chiral regime between 
the fit involving all $M_{ps}$ datasets and the fit involving the lightest four $M_{ps}$ 
datasets. Either fitting procedures can be seen to be consistent with our final estimate in 
the chiral limit is shown by black star.

\bef[hbt!]
\includegraphics[scale=0.5]{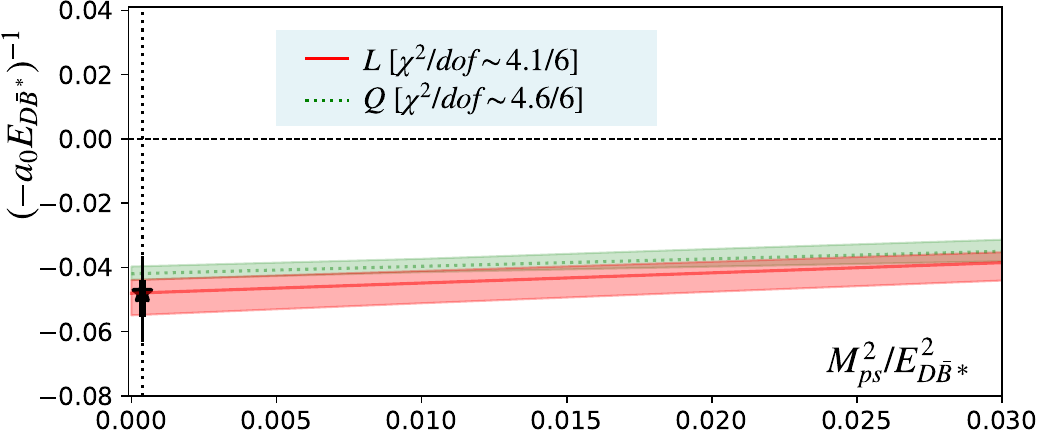}
\includegraphics[scale=0.5]{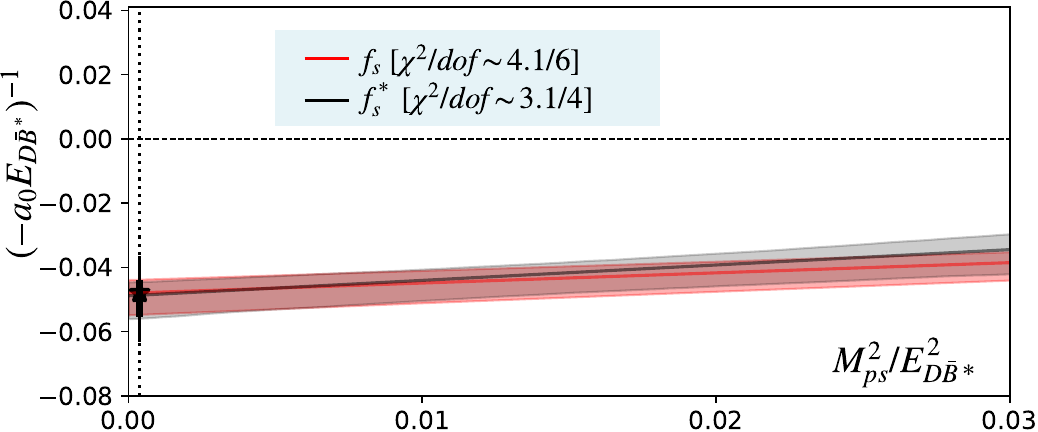}
\caption{Left: Comparison of light quark mass dependence of scattering amplitudes in the chiral regime 
determined from a linear (red band) and quadratic (green band) dependence of $k{\mathrm{cot}}\delta_0$ 
on the lattice spacing. Right: Comparison of light quark mass dependence of scattering amplitude 
$k{\mathrm{cot}}\delta_0$ in the chiral regime determined using the results from all five light quark 
masses (red band) and the results from four light light quark masses (black). The results in the physical 
limit in either cases can be seen to be consistent with the main result indicated by the star symbol. }
\eef{a0_mpi2_fits_LQ}